\documentclass[prd,twocolumn,showpacs]{revtex4}
\usepackage{graphicx}
\usepackage{psfrag}

\newcommand{ \be}{\begin{equation}}
\newcommand{ \ee}{\end{equation}}
\newcommand{ \bea}{\begin{eqnarray}}
\newcommand{ \eea}{\end{eqnarray}}
\newcommand{ \mysmall}[1]{\scriptscriptstyle #1} 
\newcommand{ \amu}{a_{\mu}}
\newcommand{ \mw}{M_{\mysmall{W}}}
\newcommand{ \mz}{M_{\mysmall{Z}}}
\newcommand{ \mh}{M_{\mysmall{H}}}
\newcommand{ \mhUB}{M_{\mysmall{H}}^{\mysmall 95}}
\newcommand{ \mhLB}{M_{\mysmall{H}}^{\mysmall \rm LB}}
\newcommand{ \mt}{M_{t}}
\newcommand{ \mpi}{m_{\pi}}
\newcommand{ \seff}{\sin^2 \!\theta_{\rm eff}^{\rm lept}}
\newcommand{ \eq}[1]{Eq.~(\ref{eq:#1})}

\newcommand{ \mev}  {\mbox{ MeV}}
\newcommand{ \bm}   {\boldmath}
\newcommand{ \ubm}  {\unboldmath}
\newcommand{ \gmt}   {$g$$-$$2$~}

\psfrag{sqrtsmev}{\sf $\sqrt s_0$~~(MeV)}
\psfrag{mhgev}{\sf \hspace{-6mm} $\mhUB$~~(GeV)}
\psfrag{leplbleplble}{\sf ~LEP LB}
\psfrag{ewubewub}{\sf EW UB}
\psfrag{dadb105}
{\hspace{-12mm}$[\Delta \alpha_{\rm had}^{(5)}(\mz)+\Delta b] \times 10^5$}
\psfrag{epspercent}{\sf $\epsilon$~~(\%)}
\psfrag{delta}{\sf $\delta$~~(MeV)}

\begin{document}

\title{The muon g-2 and the bounds on the Higgs boson mass}


\author{M.~Passera} 
\affiliation{Istituto Nazionale Fisica Nucleare,
Sezione di Padova, I-35131, Padova, Italy}
\email{massimo.passera@pd.infn.it}

\author{W.J.~Marciano}
\affiliation{Brookhaven National Laboratory, Upton, New York 11973, USA}
\email{marciano@bnl.gov}

\author{A.~Sirlin}
\affiliation{Department of Physics, New York University, 
10003 New York NY, USA}
\email{alberto.sirlin@nyu.edu}

\begin{abstract} 
\noindent 
After a brief review of the muon \gmt status, we analyze the
possibility that the present discrepancy between experiment and the
Standard Model (SM) prediction may be due to hypothetical errors in
the determination of the hadronic leading-order contribution to the
latter. In particular, we show how an increase of the hadro-production
cross section in low-energy $e^+e^-$ collisions could bridge the muon
\gmt discrepancy, leading however to a decrease on the electroweak
upper bound on $\mh$, the SM Higgs boson mass. That bound is currently
$\mh \lesssim 150$~GeV (95\%CL) based on the preliminary top quark
mass $\mt=172.6(1.4)$~GeV and the recent determination $\Delta
\alpha_{\rm had}^{(5)}(\mz) = 0.02768(22)$, while the direct-search
lower bound is $\mh > 114.4$~GeV (95\%CL). By means of a detailed
analysis we conclude that this solution of the muon \gmt discrepancy
is unlikely in view of current experimental error estimates. However,
if this turns out to be the solution, the 95\%CL upper bound on $\mh$
is reduced to about 130~GeV which, in conjunction with the
experimental lower bound, leaves a narrow window for the mass of this
fundamental particle.
\end{abstract} 


\maketitle


\subsection{Introduction}
\label{sec:INTRO}

The measurement of the anomalous magnetic moment of the muon $a_{\mu}$
by the E821 experiment at Brookhaven, with a remarkable relative
precision of 0.5 parts per million~\cite{bnl}, is challenging the
Standard Model ({\small SM}) of particle physics. Indeed, as each
sector of the {\small SM} contributes in a significant way to the
theoretical prediction of $a_{\mu} = (g-2)/2$ ($g$ is the muon's
gyromagnetic factor), this measurement allows us to test the entire
{\small SM} and provides a powerful tool to scrutinize viable ``new
physics'' appendages to this theory~\cite{NP}.

The {\small SM} prediction of the muon \gmt is conveniently split into
{\small QED}, electroweak ({\small EW}) and hadronic (leading- and
higher-order) contributions:
$
    a_{\mu}^{\mysmall \rm SM} = 
         a_{\mu}^{\mysmall \rm QED} +
         a_{\mu}^{\mysmall \rm EW}  +
         a_{\mu}^{\mbox{$\scriptscriptstyle{\rm HLO}$}} +
         a_{\mu}^{\mbox{$\scriptscriptstyle{\rm HHO}$}}.
$  
The hadronic contributions dominate the present $a_{\mu}^{\mysmall \rm
SM}$ uncertainty. The {\small QED} prediction, computed up to four
(and estimated at five) loops, currently stands at
$a_{\mu}^{\mysmall \rm QED} = 116584718.10(16)
\times 10^{-11}$\cite{QED,MPrev}, 
while the {\small EW} effects, suppressed by a factor
$(m_{\mu}/\mw)^2$, provide
$a_{\mu}^{\mysmall \rm EW} = 154(2) \times 10^{-11}$\cite{EW}.
The most recent calculations of the hadronic leading-order
contribution via the hadronic $e^+ e^-$ annihilation data, to be
discussed later, are in very good agreement:
$ a_{\mu}^{\mbox{$\scriptscriptstyle{\rm HLO}$}} = 
 6909(44) \times 10^{-11}$\cite{DE07},
$6894(46) \times 10^{-11}$\cite{HMNT06},
$6921(56) \times 10^{-11}$\cite{Jeger06}, and
$6944(49) \times 10^{-11}$\cite{TY05}. 
The higher-order hadronic term is further divided into two parts:
$
     \amu^{\mbox{$\scriptscriptstyle{\rm HHO}$}}=
     \amu^{\mbox{$\scriptscriptstyle{\rm HHO}$}}(\mbox{vp})+
     \amu^{\mbox{$\scriptscriptstyle{\rm HHO}$}}(\mbox{lbl}).
$
The first one, 
$-98\, (1) \times 10^{-11}$\cite{HMNT06},
is the $O(\alpha^3)$ contribution of diagrams containing hadronic
vacuum polarization insertions~\cite{Kr96}. The second term, also of
$O(\alpha^3)$, is the hadronic light-by-light contribution; as it
cannot be determined from data, its evaluation relies on specific
models. Recent determinations of this term vary between
$80(40) \times 10^{-11}$\cite{Andreas}
and
$136(25) \times 10^{-11}$\cite{Arkady}.
The most recent one,
$110(40) \times 10^{-11}$\cite{BP07},
lies between them. If we add this result to the leading-order hadronic
contribution, for example the value of Ref.~\cite{HMNT06} (which also
provides a recent calculation of the hadronic contribution to the
effective fine-structure constant, later required for our analysis),
and the rest of the {\small SM} contributions, we obtain
$     \amu^{\mbox{$\scriptscriptstyle{\rm SM}$}}= 116591778(61)  
\times 10^{-11}$.
The difference with the experimental value
$
    a_{\mu}^{\mbox{$\scriptscriptstyle{\rm EXP}$}}  =
               116592080(63) \times 10^{-11}$~\cite{bnl}
is 
$\Delta a_{\mu} = a_{\mu}^{\mbox{$\scriptscriptstyle{\rm EXP}$}}-
\amu^{\mbox{$\scriptscriptstyle{\rm SM}$}} = +302(88) \times 10^{-11}$,
i.e., 3.4 standard deviations (all errors were added in quadrature).
Similar discrepancies are obtained employing the values of the
leading-order hadronic contribution reported in
Refs.~\cite{DE07,Jeger06,TY05}.

The term $a_{\mu}^{\mbox{$\scriptscriptstyle{\rm HLO}$}}$ can
alternatively be computed incorporating hadronic $\tau$-decay data,
related to those of hadroproduction in $e^+e^-$ collisions via isospin
symmetry~\cite{ADH98,DEHZ}.  Unfortunately there is a large difference
between the $e^+e^-$- and $\tau$-based determinations of
$a_{\mu}^{\mbox{$\scriptscriptstyle{\rm HLO}$}}$, even if isospin
violation corrections are taken into account~\cite{IVC1}. The
$\tau$-based value is significantly higher, leading to a small ($\sim
1 \sigma$) $\Delta a_{\mu}$ difference. As the $e^+e^-$ data are more
directly related to the $\amu^{\mbox{$\scriptscriptstyle{\rm HLO}$}}$
calculation than the $\tau$ ones, the latest analyses do not include
the latter. Also, we note that recently studied additional
isospin-breaking corrections somewhat reduce the difference between
these two sets of data (lowering the $\tau$-based
determination)~\cite{IVC2,IVC3}, and a new analysis of the pion form
factor claims that the $\tau$ and $e^+e^-$ data are consistent after
isospin violation effects and vector meson mixings are
considered~\cite{IVC4}.  Recent reviews of the muon \gmt can be found
in Refs.~\cite{MPrev, DM04, Reviews}.

The 3.4 $\sigma$ discrepancy between the theoretical prediction and
the experimental value of the muon \gmt can be explained in several
ways. It could be due, at least in part, to an error in the
determination of the hadronic light-by-light contribution. However, if
this were the only cause of the discrepancy, $a_{\mu}^{\mysmall \rm
  HHO}(\mbox{lbl})$ would have to move up by many standard
deviations to fix it --~roughly eight, if we use the
$a_{\mu}^{\mysmall \rm HHO}(\mbox{lbl})$ result of Ref.~\cite{BP07}
(which includes all known uncertainties), and more than ten if the
less conservative estimate of Ref.~\cite{Arkady} is employed instead.
Although the errors assigned to $a_{\mu}^{\mysmall \rm
  HHO}(\mbox{lbl})$ are only educated guesses, this solution seems
unlikely, at least as the dominant one.

Another possibility is to explain the discrepancy $\Delta a_{\mu}$ via
the {\small QED}, {\small EW} and hadronic higher-order vacuum
polarization contributions; this looks very improbable, as one can
immediately conclude inspecting their values and uncertainties
reported above. If we assume that the \gmt experiment {\small E821} is
correct, we are left with two options: possible contributions of
physics beyond the {\small SM}, or an erroneous determination of the
leading-order hadronic contribution
$a_{\mu}^{\mbox{$\scriptscriptstyle{\rm HLO}$}}$ (or combinations of
the two). The first of these two options has been widely discussed in
the literature; we will focus on the second one, and analyze its
implications for the {\small EW} bounds on the mass of the Higgs
boson.

\subsection{Shifts of  
\bm $a_{\mu}^{\mbox{$\scriptscriptstyle{\rm HLO}$}}$ and  
$\Delta \alpha_{\rm had}^{(5)}(\mz)$ \ubm }

The evaluation of the hadronic leading-order contribution
$\amu^{\mbox{$\scriptscriptstyle{\rm HLO}$}}$, due to the hadronic
vacuum polarization correction to the one-loop {\small QED} diagram,
involves long-distance {\small QCD} for which perturbation theory
cannot be employed. However, using analyticity and unitarity, it was
shown long ago that this term can be computed from hadronic $e^+ e^-$
annihilation data via the dispersion integral~\cite{DISPamu}
\be
      a_{\mu}^{\mbox{$\scriptscriptstyle{\rm HLO}$}}= 
      \frac{1}{4\pi^3} \!
      \int^{\infty}_{4m_{\pi}^2} ds \, K(s) \, \sigma (s),
\label{eq:amudispint}
\ee
where $\sigma (s)$ is the total cross section for $e^+ e^-$
annihilation into any hadronic state, with extraneous {\small QED}
corrections subtracted off, and $s$ is the squared momentum
transfer. The kernel $K(s)$ is the well-known function
\be
                K(s)= \int_0^1 \!dx \frac{x^2 (1-x)}
                {x^2 +(1-x)s/m_\mu^2}
\ee
(see Ref.~\cite{EJ95} for some of its explicit representations and
their suitability for numerical evaluations). It decreases
monotonically for increasing $s$ and, for large $s$, it behaves as
$m_\mu^2/(3s)$ to a good approximation. One finds that the low-energy
region of the dispersive integral is enhanced by $\sim 1/s^2$. About
90\% of the total contribution to $\amu^{\mbox{$\scriptscriptstyle{\rm
      HLO}$}}$ is accumulated at center-of-mass energies $\sqrt{s}$
below 1.8~GeV and roughly three-fourths of
$\amu^{\mbox{$\scriptscriptstyle{\rm HLO}$}}$ is covered by the
two-pion final state which is dominated by the $\rho(770)$
resonance~\cite{DEHZ}. Note that $\amu^{\mbox{$\scriptscriptstyle{\rm
      HLO}$}}$ is a positive definite quantity. Exclusive low-energy
$e^+e^-$ cross sections have been measured by experiments running at
$e^+e^-$ colliders in Frascati, Novosibirsk, Orsay, and Stanford,
while at higher energies the total cross section has been measured
inclusively. Perturbative {\small QCD} becomes applicable at higher
loop-momenta, so that at some energy scale one can switch from data to
{\small QCD}~\cite{pQCD}.

Let's now assume that the discrepancy
$\Delta a_{\mu} = a_{\mu}^{\mbox{$\scriptscriptstyle{\rm EXP}$}}-
\amu^{\mbox{$\scriptscriptstyle{\rm SM}$}} = +302(88) \times 10^{-11}$,
is due to --~and only to~-- hypothetical mistakes in $\sigma (s)$, and
let us increase this cross section in order to raise
$\amu^{\mbox{$\scriptscriptstyle{\rm HLO}$}}$, thus reducing $\Delta
a_{\mu}$. This simple assumption leads to interesting consequences. An
upward shift of the hadronic cross section also induces an increase of
the value of the hadronic contribution to the effective fine-structure
constant at $M_Z$~\cite{DISPDalpha},
\be
 \Delta \alpha_{\rm had}^{(5)}(\mz) = \frac{\mz^2}{4 \alpha \pi^2}
  \,\, P \! \int_{4m_\pi^2}^{\infty} ds \, \frac{\sigma(s)}{\mz^2 -s}
\label{eq:Dpi5dispint}
\ee       
($P$ stands for Cauchy's principal value).  This integral is similar
to the one we encountered in \eq{amudispint} for
$a_{\mu}^{\mbox{$\scriptscriptstyle{\rm HLO}$}}$. There, however, the
weight function in the integrand gives a stronger weight to low-energy
data.  The negligible contribution to $a_{\mu}^{\mysmall \rm HLO}$ and
$\Delta \alpha_{\rm had}^{(5)}(\mz)$ of the $\pi^0 \gamma$ channel
below the $\pi^{+} \pi^{-}$ threshold was ignored in
Eqs.~(\ref{eq:amudispint},\ref{eq:Dpi5dispint}).  Let us define
\bea
     a &=& \int_{4m_{\pi}^2}^{s_u}ds \, f(s) \, \sigma (s),
\label{eq:adef} \\ 
     b &=& \int_{4m_{\pi}^2}^{s_u}ds \, g(s) \, \sigma (s),
\label{eq:bdef} 
\eea
where the upper limit of integration is $s_u < \mz^2$, and the kernels
are $f(s) = K(s)/(4 \pi^3)$ and $g(s) = [\mz^2/(\mz^2-s)]/(4 \alpha
\pi^2)$. Equations (\ref{eq:adef},\ref{eq:bdef}) provide the
contributions to $a_{\mu}^{\mbox{$\scriptscriptstyle{\rm HLO}$}}$ and
$\Delta \alpha_{\rm had}^{(5)}(\mz)$, respectively, in the region from
the two-pion threshold up to $s_u$ (see
Eqs.~(\ref{eq:amudispint},\ref{eq:Dpi5dispint})).

An increase of the cross section $\sigma(s)$ of the form
\be
     \Delta \sigma(s) = \epsilon \sigma(s)
\label{eq:shifts} 
\ee
in the energy range $\sqrt s \in [\sqrt s_0 - \delta/2, \sqrt s_0 +
  \delta/2]$, where $\epsilon$ is a positive constant and
$2m_{\pi}+\delta/2<\sqrt s_0<\sqrt s_u -\delta/2$, increases $a$ by
$\Delta a (\sqrt s_0,\delta,\epsilon) = \epsilon \int_{\sqrt
  s_0-\delta/2}^{\sqrt s_0+\delta/2} 2t \, \sigma(t^2) \, f(t^2) \,
dt$. If we assume that the muon \gmt discrepancy is entirely due to
this increase in $\sigma(s)$ so that $\Delta a (\sqrt
s_0,\delta,\epsilon) = \Delta a_{\mu}$, the parameter $\epsilon$
becomes
\be
       \epsilon =  
       \frac{\Delta a_{\mu}}{
         \int_{\sqrt s_0-\delta/2}^{\sqrt s_0 +\delta/2} 
         2t \, f(t^2) \, \sigma(t^2) \, dt},
\label{eq:eps} 
\ee
and the corresponding increase in $\Delta \alpha_{\rm had}^{(5)}(\mz)$
is
\be 
\Delta b(\sqrt s_0,\delta) = \Delta a_{\mu} 
\frac{\int_{\sqrt s_0-\delta/2}^{\sqrt s_0+\delta/2} g(t^2) \, 
  \sigma(t^2) \, t \, dt} 
     {\int_{\sqrt s_0-\delta/2}^{\sqrt s_0+\delta/2} f(t^2) \, 
  \sigma(t^2) \, t \, dt}.
\label{eq:shiftb} 
\ee
In the limiting case of a point-like shift 
$\Delta \sigma(s) \!=\! \epsilon' \delta(s-s_0)$, 
with $2m_{\pi} < \sqrt s_0 < \sqrt s_u$, the condition
$\Delta a (\sqrt s_0,\epsilon') = \Delta a_{\mu}$, 
with
$\Delta a (\sqrt s_0,\epsilon') = \epsilon' f(s_0)$, 
leads to
\be
\Delta b(\sqrt s_0) = \Delta a_{\mu} \left[g(s_0)/f(s_0)\right].
\label{eq:shiftb_simple} 
\ee
Following Ref.~\cite{DEHZ}, to overcome the lack of precise data for
$\sigma(s)$ at threshold energies, in the region $2\mpi < \sqrt s <
500 \mev$ one can adopt the polynomial parametrization for the pion
form factor $F_{\pi}(s)$ inspired by chiral perturbation theory; the
parameters are determined from a fit to the data for both time-like
and space-like momentum transfers~\cite{EJ95,DEHZ,Co04}. The cross
section below 500 MeV is therefore given by
\be
\sigma(s) = \frac{\pi \alpha^2}{3s} \beta_{\pi}^3 \, |F_{\pi}(s)|^2,
\label{eq:sigmalow}
\ee 
where $\beta_{\pi} = (1-4\mpi^2/s)^{1/2}$, $F_{\pi}(s) = 1 + s \langle
r^2\rangle_{\pi}/6 + s^2 c_1 +s^3 c_2$, $\langle r^2\rangle_{\pi} =
(0.439 \pm 0.008)~\mbox{fm}^2$, $c_1 = (6.8 \pm 1.9)~\mbox{GeV}^{-4}$,
and $c_2 = (-0.7 \pm 6.8)~\mbox{GeV}^{-6}$ (see Ref.~\cite{DEHZ} for
the correlation matrix of these coefficients). Between 500~MeV and
1.4~GeV we use the cross section directly obtained combining the
experimental results of the
$\pi^+ \pi^-$~\cite{2pi}, 
$\pi^+ \pi^-\pi^0$~\cite{3pia,3pib2K}, 
$K^+K^-$~\cite{3pib2K,2Kc}, 
$K^0_L K^0_S$~\cite{2K0}, 
$2\pi^+ 2\pi^-$~\cite{4pic},
$\pi^0 \pi^0 \pi^+ \pi^-$~\cite{4pi0},
$\pi^0 \gamma$~\cite{1pi,1et} and
$\eta \gamma$~\cite{1et}
channels. Between 1.4~GeV and 2~GeV we employ the inclusive
measurements of Refs.~\cite{inclusive}.

\begin{figure}
\includegraphics[width=85mm,angle=0]{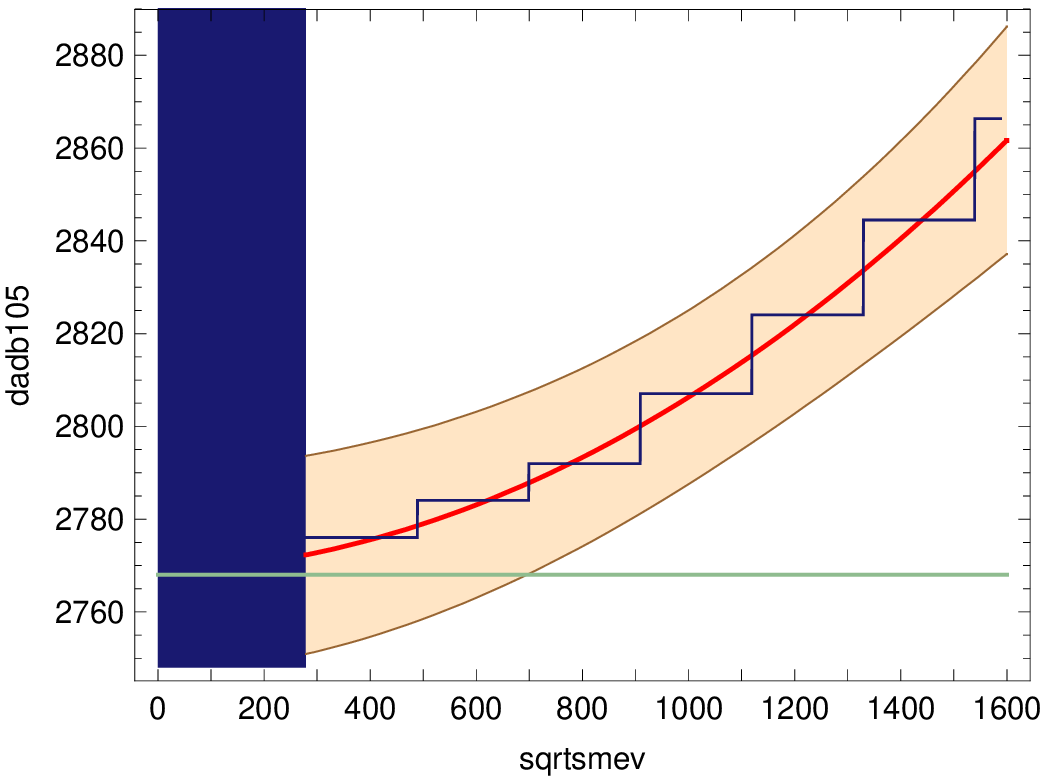}
\vspace{0cm}\caption{Shifts of $\Delta \alpha_{\rm had}^{(5)}(\mz)$.
  The histogram indicates the increase $\Delta b(\sqrt s_0,\delta)$
  obtained varying the cross section by $\Delta \sigma(s) = \epsilon
  \sigma(s)$ in $\delta\!=\!210$~MeV energy regions, while $\Delta
  b(\sqrt s_0)$, obtained for point-like increases, is plotted as a
  smooth curve. The shifts are added to $\Delta \alpha_{\rm
    had}^{(5)}(\mz) = 0.02768(22)$~\cite{HMNT06} (horizontal line).
  The uncertainty of the sum $\Delta \alpha_{\rm had}^{(5)}(\mz) +
  \Delta b(\sqrt s_0)$ is shown by the light band.}
\label{fig:dalpha}
\end{figure}

Figure 1 shows the shifts $\Delta b(\sqrt s_0,\delta\!=\!210\mbox{MeV})$
(histogram) and $\Delta b(\sqrt s_0)$ (smooth curve) obtained from the
increases $\Delta \sigma(s) = \epsilon \sigma(s)$ and $\Delta
\sigma(s) = \epsilon' \delta(s-s_0)$, respectively. These shifts, shown
as functions of $\sqrt s_0$, are added to the value $\Delta
\alpha_{\rm had}^{(5)}(\mz) = 0.02768(22)$~\cite{HMNT06}. The
uncertainty of the sum $\Delta \alpha_{\rm had}^{(5)}(\mz) + \Delta b
(\sqrt s_0)$ is indicated by the light band.
To compute it, we first note that the errors $46 \times 10^{-11}$ in
$a_{\mu}^{\mbox{$\scriptscriptstyle{\rm HLO}$}}$ and $22 \times
10^{-5}$ in $\Delta \alpha_{\rm had}^{(5)}(\mz)$~\cite{HMNT06} 
are strongly correlated since they arise mainly from the same source,
namely the uncertainty in the hadronic
$e^+ e^-$ annihilation cross section (which includes the uncertainties
associated with the radiative corrections applied to the experimental
data). Taking this into account, and observing also that the error in
$\Delta b(\sqrt s_0)$ due to the
$a_{\mu}^{\mbox{$\scriptscriptstyle{\rm HLO}$}}$ uncertainty is $-46
\times 10^{-11}[g(s_0)/f(s_0)]$, we add it linearly to $22 \times
10^{-5}$, and then combine in quadrature this result with the error in
$\Delta b(\sqrt s_0)$ induced by the remaining $\Delta a_{\mu}$
uncertainty. (We note that combining all errors in quadrature,
ignoring their correlation, would enlarge the uncertainty of the sum
$\Delta \alpha_{\rm had}^{(5)}(\mz) + \Delta b (\sqrt s_0)$, but would
only induce minimal changes in our analysis.) The uncertainty of the
sum $\Delta \alpha_{\rm had}^{(5)}(\mz) + \Delta b (\sqrt s_0,
\delta)$, for finite energy intervals, is computed analogously,
neglecting the relative error of the ratio of integrals on the
r.h.s.\ of \eq{shiftb} with respect to the large relative error of
$\Delta a_{\mu}$.  The dark area below $2\mpi$, where $\mpi$ is the
mass of the charged pion, denotes the kinematically forbidden region
below the $\pi^{+} \pi^{-}$ threshold (the $\pi^0 \gamma$ channel is
neglected below this threshold).
%

\subsection{Connection with the Higgs boson mass}

The dependence of {\small SM} predictions, via quantum effects, on the
mass of the Higgs boson $\mh$ provides a powerful tool to set indirect
bounds on the mass of this fundamental missing piece of the {\small
SM}.  Indeed, comparing calculated quantities with their precise
experimental values, the present global fit of the {\small LEP}
Electroweak Working Group ({\small LEP-EWWG}) leads to the value
$\mh = 87^{+36}_{-27}$~GeV 
and to the 95\% confidence level ({\small CL}) upper bound $\mhUB
\simeq 160$~GeV~\cite{newLEPEWWG}. This result is based on the very
recent preliminary top quark mass $\mt=172.6(1.4)$~GeV from a
combined CDF-D0 fit~\cite{Group:2008nq} and the value 
$\Delta \alpha_{\rm had}^{(5)}(\mz) = 0.02758(35)$~\cite{BP05}.
The {\small LEP} direct-search lower bound is
$\mhLB=114.4$~GeV~\cite{MHLB03}, also at the 95\% {\small CL}.

Although the global fit to the {\small EW} data employs a large set of
observables, the $\mh$ upper bound is strongly driven by the
comparison of the theoretical predictions of the mass of the W boson
and the effective {\small EW} mixing angle $\seff$ with their
precisely measured values~\cite{FOS}. Convenient formulae providing
the {\small SM} prediction of $\mw$ and $\seff$ in terms of $\mh$, the
top quark mass $\mt$, $\Delta \alpha_{\rm had}^{(5)}(\mz)$, and
$\alpha_s(\mz)$, the value of the strong coupling constant at the
scale $\mz$, are given in Refs.~\cite{formulette}.  Combining these
$\mw$ and $\seff$ predictions by means of a numerical
$\chi^2$-analysis, and using the present world-average values
$\mw = 80.398(25)$~GeV~\cite{LEPEWWG06,Wmass, Gru07}, 
$\seff = 0.23153(16)$~\cite{LEPEWWG05},
$\mt = 172.6(1.4)$~GeV~\cite{Group:2008nq}, 
$\alpha_s(\mz) = 0.118(2)$~\cite{PDG06},
and the determination
$\Delta \alpha_{\rm had}^{(5)}(\mz) = 0.02758(35)$~\cite{BP05}
adopted by the {\small LEP-EWWG}, we obtain
$\mh = 92^{+38}_{-28}$~GeV
and $\mhUB=161$~GeV. We see that indeed the $\mh$ values obtained from
the $\mw$ and $\seff$ predictions are quite close to the results of
the global analysis.

The $\mh$ dependence of the {\small SM} prediction of the muon
$g$$-$$2$, via its {\small EW} contribution, is too weak to provide
$\mh$ bounds from the comparison with the measured value. Indeed, the
shift of $a_{\mu}^{\mysmall \rm SM}$ for $\mh$ varying between
114.4~GeV and 300~GeV is only of $O(10^{-11})$, which is negligible
when compared with the hadronic and experimental uncertainties.
On the other hand, $\Delta \alpha_{\rm had}^{(5)}(\mz)$ is one of the
key inputs of the {\small EW} fits. For example, employing the recent
(slightly higher) value
$\Delta \alpha_{\rm had}^{(5)}(\mz) = 0.02768(22)$~\cite{HMNT06}
instead of 
$\Delta \alpha_{\rm had}^{(5)}(\mz) = 0.02758(35)$~\cite{BP05},
the $\mh$ prediction shifts down to
$\mh = 90^{+33}_{-25}$~GeV 
and $\mhUB=150$~GeV. We note that $\mhUB$ depends both on the central
value and on the uncertainty of $\Delta \alpha_{\rm
  had}^{(5)}(\mz)$. Henceforth, we employ the recent evaluation
$\Delta \alpha_{\rm had}^{(5)}(\mz) = 0.02768(22)$~\cite{HMNT06}. (For
the dependence of $\mh$ and its bounds on $\Delta \alpha_{\rm
  had}^{(5)}(\mz)$ see Refs.~\cite{formulette}).  Next we consider the
new values of $\Delta \alpha_{\rm had}^{(5)}(\mz)$ obtained shifting
0.02768(22) by $\Delta b(\sqrt s_0)$ and $\Delta b(\sqrt s_0, \delta)$
(including their uncertainties as discussed in the previous section),
and compute the corresponding new values of $\mhUB$ by means of the
combined $\chi^2$-analysis based on the $\mw$ and $\seff$ inputs. The
results are shown in Fig.\ 2. The lower region $\mh < 114.4$~GeV is
excluded by the direct {\small LEP} searches at $95\%$ {\small CL},
while the upper one is excluded by the indirect {\small EW} $95\%$
{\small CL} bound $\mh < 150$~GeV obtained with $\Delta \alpha_{\rm
  had}^{(5)}(\mz) = 0.02768(22)$. (As in the case of $\Delta
\alpha_{\rm had}^{(5)}(\mz)$, the value adopted here for
$\amu^{\mbox{$\scriptscriptstyle{\rm HLO}$}}$ is from the recent
article in Ref.~\cite{HMNT06}.) If we increase the hadronic cross
section $\sigma(s)$ by $\epsilon' \delta(s-s_0)$ in order to bridge
the muon \gmt discrepancy $\Delta a_{\mu}$, $\mhUB$ decreases, as
shown by the continuous red line in Fig.\ 2, further restricting the
already narrow allowed region for $\mh$. In particular, this curve
falls below $\mhLB$ for $\sqrt s_0 \gtrsim 1.1$~GeV. The two
histograms show the $\mhUB$ values when the analysis is repeated with
$\Delta \sigma = \epsilon \sigma (s)$
shifts in $\delta=210$~MeV and $\delta=400$~MeV energy regions. We
conclude that the hypothetical shifts
$\Delta \sigma = \epsilon \sigma (s)$
(in $\sqrt s \in [\sqrt s_0 - \delta/2, \sqrt s_0 + \delta/2]$) of the
hadronic cross section that bridge the muon \gmt discrepancy,
conflict with the {\small LEP} lower limit when
$\sqrt s_0 > (\sqrt s_0)_{\rm thr} \sim 1.2$~GeV, 
for values of $\delta$ up to several hundreds of MeV. The threshold
$(\sqrt s_0)_{\rm thr}$ increases above $\sim 1.3$~GeV for
hypothetical shifts $\epsilon \sigma (s)$ in even wider energy regions
$\delta \gtrsim 1$~GeV, but uniform shifts of the cross section in
such wide energy ranges appear to be unrealistic.
\begin{figure}
\includegraphics[width=85mm,angle=0]{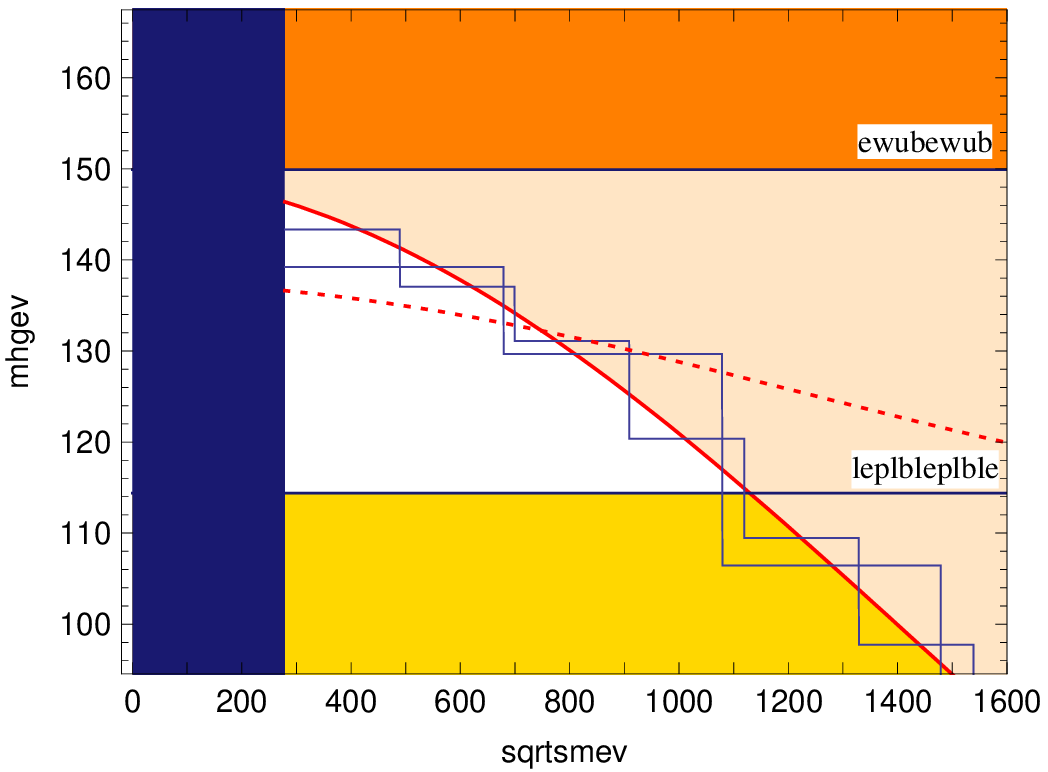}
\vspace{0cm}\caption{The $\mhUB$ values obtained via the $\mw$ and
  $\seff$ fits using as input for $\Delta \alpha_{\rm had}^{(5)}(\mz)$
  the value 0.02768(22) increased by $\Delta b(\sqrt s_0)$ (smooth
  curve) and by $\Delta b(\sqrt s_0,\delta\!=\!210\mbox{MeV},
  400\mbox{MeV})$ (histograms). The area below 114.4~GeV, partly
  yellow and partly pink, is excluded at $95\%$ {\small CL} by the
  {\small LEP} direct lower bound, while the orange $\mh>150$~GeV one
  is forbidden by the {\small EW} indirect upper bound. As in Fig.~1,
  the region $\sqrt s_0 < 2m_{\pi}$ is excluded. The dotted line
  replaces the smooth one when $\tau$ data are incorporated in the
  determination of $\Delta \alpha_{\rm had}^{(5)}(\mz)$ and
  $a_{\mu}^{\mysmall \rm SM}$.}
\label{fig:mhtot}
\end{figure}

If $\tau$ data are incorporated in the calculation of the dispersive
integrals in Eqs.~(\ref{eq:amudispint}) and (\ref{eq:Dpi5dispint}),
the leading-order hadronic contribution to the muon \gmt significantly
increases to
$ a_{\mu}^{\mbox{$\scriptscriptstyle{\rm HLO}$}} = 
 7110(58) \times 10^{-11}$\cite{DEHZ}, 
the higher-order vacuum polarization term slightly decreases to      
$\amu^{\mbox{$\scriptscriptstyle{\rm HHO}$}}(\mbox{vp}) = -101(1)
\times 10^{-11}$\cite{HMNT06,DM04}, and the discrepancy with the
experimental value drops to
$\Delta a_{\mu} = +89(95) \times 10^{-11}$,
i.e.\ roughly 1 $\sigma$. While using $\tau$ data almost solves the
muon \gmt discrepancy, it increases the value of $\Delta \alpha_{\rm
  had}^{(5)}(\mz)$ to 0.02782(16)~\cite{Marciano04,DEHZ}. In
Ref.~\cite{Marciano04} it was shown that this increase leads to a low
$\mh$ prediction which is suggestive of a near conflict with $\mhLB$,
leaving a very narrow window for $\mh$.  Indeed, with this value of
$\Delta \alpha_{\rm had}^{(5)}(\mz)$ and the same above-discussed
values of the other inputs of the $\chi^2$-analysis, we find
$\mh = 84^{+30}_{-23}$~GeV
and an $\mhUB$ value of only 138~GeV. The dotted line in Fig.~2 shows
the $\mhUB$ values obtained using $\tau$ data to compute $\Delta
\alpha_{\rm had}^{(5)}(\mz)$ and $\Delta a_{\mu}$, with the hadronic
cross section $\sigma(s)$ increased by $\epsilon' \delta(s-s_0)$ in
order to bridge the $\Delta a_{\mu}$ difference.

As we briefly mentioned in the Introduction, recently computed
isospin-breaking violations, further improvements of the long-distance
radiative corrections to the decay $\tau^- \to \pi^- \pi^0
\nu_{\tau}$~\cite{IVC2} and differentiation of the neutral and charged
$\rho$ properties~\cite{IVC3}, reduce to some extent the difference
between $\tau$ and $e^+e^-$ data, lowering the $\tau$-based
determination of $ a_{\mu}^{\mbox{$\scriptscriptstyle{\rm
      HLO}$}}$. Moreover, a recent analysis of the pion form factor
below 1~GeV claims that $\tau$ data are consistent with the $e^+e^-$
ones after isospin violation effects and vector meson mixings are
considered~\cite{IVC4}. In this case one could therefore use the
$e^+e^-$ data below $\sim 1$~GeV, confirmed by the $\tau$ ones, and
assume that $\Delta a_{\mu}$ is accommodated by hypothetical errors
occurring above $\sim 1$~GeV, where disagreement persists between
these two sets of data. Our previous analysis shows that this
assumption would lead to values of $\mhUB$ inconsistent with the
{\small LEP} lower bound.

It is interesting to note that there are more complex scenarios where
it is possible to bridge the $\Delta a_{\mu}$ discrepancy without
significantly affecting $\mhUB$. For instance, we may envisage an
increase of $\sigma(s)$ at low $s$ combined with a decrease at high
$s$ in such a manner that their overall contribution to $\Delta
\alpha_{\rm had}^{(5)}(\mz)$, and therefore to $\mhUB$, approximately
cancels. Since the contributions to
$a_{\mu}^{\mbox{$\scriptscriptstyle{\rm HLO}$}}$ are more heavily
weighted at low $s$, it is then possible to further adjust the
positive and negative $\sigma(s)$ shifts to bridge the muon \gmt
discrepancy. However, such scheme requires two fine-tuning steps and a
larger increase of $\sigma(s)$ at low $s$, and is therefore
considerably more unlikely than the simplest scenarios, involving a
single adjustable contribution, that are discussed in detail in this
paper.

\subsection{How realistic are these shifts \bm  $\Delta \sigma(s)$\ubm?}
\label{sec:REALISTIC}

In the above study, the hadronic cross section $\sigma(s)$ was shifted
up by amounts required to adjust the muon \gmt discrepancy $\Delta
a_{\mu}$. Apart from the implications for the Higgs boson mass (and
the restrictions deriving from them), these shifts may actually be
inadmissibly large when compared with the quoted experimental
uncertainties. For example, one of the histograms in Fig.~2 shows that
a shift $\Delta \sigma$ in a 210~MeV bin centered just above the
$\rho$ peak could fix the muon \gmt discrepancy (lowering $\mhUB$ to
131~GeV); but is such a shift of the precisely measured cross section
at the $\rho$ peak realistic?

To investigate this problem, we turn our attention to the parameter
$\epsilon=\Delta \sigma(s)/\sigma(s)$, i.e.\ the ratio of the shift
$\Delta \sigma(s)$ required to bridge the muon \gmt discrepancy and
the cross section $\sigma(s)$, provided by \eq{eps}. Clearly, the
value of $\epsilon$ depends on the choice of the energy range $[\sqrt
  s_0 - \delta/2, \sqrt s_0 + \delta/2]$ where $\sigma(s)$ is
increased and, for fixed $\sqrt s_0$, it increases when $\delta$
decreases. The minimum value of $\epsilon$ is roughly $+4\%$; it
occurs if the hadronic cross section $\sigma(s)$ is multiplied by
$(1+\epsilon)$ in the whole integration region of \eq{amudispint},
from the $\pi^{+} \pi^{-}$ threshold to infinity (this minimum value
of $\epsilon$ changes only negligibly whether the shift up of
$\sigma(s)$ includes or not the high-energy region where perturbative
{\small QCD} is employed). Such a shift would lead to an $\mhUB$ of
roughly 75~GeV, well below the {\small LEP} lower bound.

\begin{figure}[h]
\includegraphics[width=88mm,angle=0]{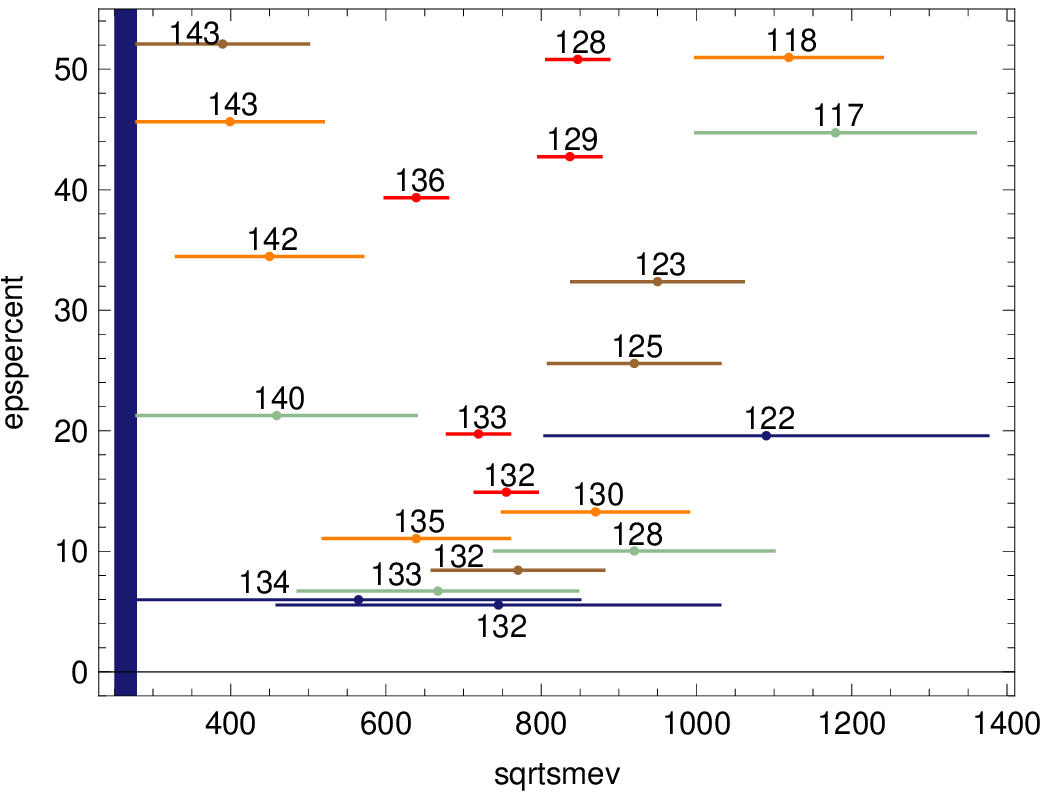}
\vspace{0cm}\caption{Values of $\epsilon$ obtained increasing
  $\sigma(s)$ by $\epsilon \sigma(s)$, to bridge the $\Delta a_{\mu}$
  discrepancy, in energy ranges $[\sqrt s_0 - \delta/2, \sqrt s_0 +
    \delta/2]$ for various values of $\sqrt s_0$ and $\delta$. The
  number next to each segment indicates the $\mhUB$ value (in GeV)
  induced by the $\epsilon \sigma(s)$ shift in that energy
  region. Same length segments are of the same color. The midpoint of
  each segment is displayed by a dot.}
\label{fig:subway}
\end{figure}

Figure \ref{fig:subway} shows the values of $\epsilon$ (in per cent)
for several bin widths $\delta$ and central values $\sqrt s_0$ (same
length segments are of the same color). Also, next to each segment we
quote the value of $\mhUB$ (in GeV) obtained performing the shift
$\Delta \sigma = \epsilon \sigma (s)$ in that energy range. A shift up
of $\sigma(s)$ in the energy range from $2\mpi$ to 850~MeV, to fix
$\Delta a_{\mu}$, leads to $\epsilon \sim 6\%$ and lowers $\mhUB$ to
134~GeV. Higher values of $\epsilon$ are obtained for narrower energy
bins, particularly if they do not include the $\rho$-$\omega$
resonance region. For example, a huge $\epsilon \sim 52\%$ increase is
needed to accommodate $\Delta a_{\mu}$ with a shift of the cross
section in the region from $2\mpi$ up to 500~MeV (reducing $\mhUB$ to
143~GeV), while an increase in a bin of the same size but centered at
the $\rho$ peak requires $\epsilon \sim 8\%$ (lowering $\mhUB$ to
132~GeV). As the quoted experimental uncertainty of $\sigma(s)$ below
1~GeV is of the order of a few per cent (or less, in some specific
energy regions), the possibility to explain the muon \gmt discrepancy
with these shifts $\Delta \sigma(s)$ appears to be unlikely. Figure
\ref{fig:subway} shows that for fixed $\delta$ (i.e., segments of the
same color), lower values of $\epsilon$ are obtained if the shifts
occur in energy ranges centered around the $\rho$-$\omega$ resonances;
but also this possibility looks unlikely, since it requires variations
of $\sigma(s)$ of at least $\sim 6$\%. If, however, we allow
variations of the cross section up to $\sim 6$\% (7\%), $\mhUB$ is
reduced to less than $\sim 134$~GeV (135~GeV). For example, the
$\sim6$\% shifts in the intervals [0.5,1.0]~GeV or [0.6,1.2]~GeV,
required to fix $\Delta a_{\mu}$ (not represented in
Fig.~\ref{fig:subway}), lower $\mhUB$ to 133~GeV or 130~GeV,
respectively.

We remind the reader that the present experimental results for $\seff$
exhibit an intriguing dichotomy.  Those based on the leptonic
observables
lead to $(\seff)_{l}=0.23113(21)$, while the average of those derived
from the hadronic sector
is $(\seff)_{h}=0.23222(27)$~\cite{LEPEWWG05}. The results within each
group agree well with each other, but the averages of the two sectors
differ by about $3.2 \sigma$. Our analysis, like the {\small LEP-EWWG}
one, depends on the value of $\seff$.  For instance, if we were to use
$(\seff)_{h}$, we would obtain a significantly higher {\small SM}
prediction:
$\mh = 129^{+53}_{-40}$~GeV,
$\mhUB = 225$~GeV, 
and a continuous (red) line in Fig.~\ref{fig:mhtot} similarly shifted
up. However, we note that in this scenario the $\mh$ predictions from
$\mw$ and $(\seff)_{h}$ are inconsistent with one another unless one
introduces additional ``new physics'' beyond the {\small SM}. For
example, the difference could be associated with a value $S \sim 0.4$
to 0.5 of the $S$-parameter, an effect generally attributed to
technicolor-like theories with additional heavy fermion chiral
doublets~\cite{Marciano:2006zu}. Instead, if we were to employ
$(\seff)_{l}$, the {\small SM} prediction would drop sharply to
$\mh = 50^{+25}_{-18}$~GeV,
$\mhUB=97$~GeV,
which is already in conflict with the direct-search lower bound. Thus,
in that case, no shift $\Delta \sigma(s)$ could reconcile the \gmt
discrepancy without violating the lower bound. In this paper we employ
as input the world-average of $\seff$ since this is the value
determined in the global analysis of the {\small SM}.

The $\mh$ upper bounds presented in this article depend sensitively on
the central value $\mt=172.6$~GeV and its uncertainty $\delta
\mt=1.4$~GeV. In the future, the former may still change and the
latter will further decrease. We therefore provide the following
simple formulae to translate easily the $\mhUB=150$~GeV result of our
numerical $\chi^2$-analysis based on the $\mw$ and $\seff$
predictions, as well as the $\mhUB[0.6,1.2] = 130$~GeV upper bound
corresponding to the $\sim 6$\% increase of $\sigma(s)$ in the
interval [0.6,1.2]~GeV (an illustrative case that accounts for $\Delta
a_{\mu}$), into the new values derived with different $\mt$ and $\delta
\mt$ inputs:
\bea
  &&\mhUB = \left( 150.5 + 11.2 x +  9.4 y \right){\rm GeV,}
\label{eq:fit1}
\\ &&\mhUB[0.6,1.2] = ( 130.7 + 9.9 x + 8.2 y){\rm \, GeV,~~~~~~}
\label{eq:fit2}
\eea
with $x=\mt- 172.6{\rm GeV}$ and $y=\delta \mt - 1.4{\rm GeV}$.  Note
that, in case of a future rise of the $\mt$ central value, the
increase induced on the $\mh$ upper bounds would be partially
compensated by a reduction of the error $\delta \mt$.  Equations
(\ref{eq:fit1},\ref{eq:fit2}) reproduce the results of the detailed
numerical $\chi^2$-analysis with maximum absolute deviations of
roughly 1~GeV when $\mt \in [171,174]$~GeV and $\delta \mt \in
[1.0,1.8]$~GeV.

\subsection{Conclusions}
\label{sec:CONC}

The present discrepancy between the {\small SM} prediction of the
anomalous magnetic moment of the muon and its experimental
determination could be due to the contribution of new, yet
undiscovered, physics beyond the {\small SM}, or to errors in the
determination of the hadronic contributions. In this letter we
considered the second hypothesis and, in particular, the possibility
to accommodate the discrepancy
$\Delta a_{\mu} = +302(88) \times 10^{-11}$ 
(3.4 $\sigma$) by changes in the hadronic cross section $\sigma(s)$
used to determine the leading hadronic contribution
$a_{\mu}^{\mbox{$\scriptscriptstyle{\rm HLO}$}}$. This option has
important consequences on $\mhUB$, the 95\% {\small CL} {\small EW}
upper bound on the mass of the {\small SM} Higgs boson.

We first analyzed the effects induced by these hypothetical changes
$\Delta \sigma(s)$ on the value of $\Delta \alpha_{\rm
  had}^{(5)}(\mz)$, one of the key inputs of the {\small EW} fits with
a strong influence on the {\small SM} $\mh$ predictions. The
comparison of the theoretical predictions of $\mw$ and the effective
{\small EW} mixing angle $\seff$ with their precisely measured values
allowed us to determine, via a combined $\chi^2$ analysis, the
variations of $\mhUB$ induced by the shifts $\Delta \sigma(s)$.  We
concluded that if the hadronic cross section is shifted up in energy
regions centered above $\sim 1.2$~GeV to bridge the muon \gmt
discrepancy, the Higgs mass upper bound becomes inconsistent with the
{\small LEP} lower limit.

If $\tau$-decay data are incorporated in the calculation of
$a_{\mu}^{\mysmall \rm SM}$, the discrepancy $\Delta a_{\mu}$ drops to
$+89(95) \times 10^{-11}$.  While this almost solves the muon \gmt
discrepancy, it raises the value of $\Delta \alpha_{\rm
  had}^{(5)}(\mz)$ leading to $\mhUB= 138$~GeV, increasing the tension
with the {\small LEP} lower bound. One could also consider a scenario,
suggested by recent studies, where the $\tau$ data confirm the
$e^+e^-$ ones below $\sim 1$~GeV, while a discrepancy between them
persists at higher energies. If, in this case, $\Delta a_{\mu}$ is
reconciled by hypothetical errors above $\sim 1$~GeV, where the data
sets disagree, one also finds values of $\mhUB$ inconsistent with the
114.4~GeV lower bound. For example, if $\sigma(s)$ is shifted in the
interval [1.0,1.8]~GeV, we obtain $\mhUB= 108$~GeV.

We then questioned the plausibility of the variations $\Delta
\sigma(s)\!=\!\epsilon \sigma(s)$ required to fix $\Delta
a_{\mu}$. Their amounts clearly depend on the energy regions chosen
for the change, but we showed that they are generally very large when
compared with the actual experimental uncertainties.  Given the small
experimental uncertainty of $\sigma(s)$ below 1~GeV, the possibility
to bridge the muon \gmt discrepancy with shifts of the hadronic cross
section appears to be unlikely. Smaller values of $\epsilon$ (for
fixed bin-widths $\delta$) are needed when the shifts occur in energy
regions centered around the $\rho$-$\omega$ resonances; but also this
possibility looks unlikely since it requires variations of $\sigma(s)$
of at least $\sim 6$\%, a large modification given current
experimental error estimates. However, if this turns out to be the
solution of the $\Delta a_{\mu}$ discrepancy, we conclude that $\mhUB$
is reduced to roughly 130~GeV which, in conjunction with the 114.4~GeV
lower bound, leaves a narrow window for the mass of this fundamental
particle. Simple formulae were also provided to translate $\mh$ upper
bounds derived in this paper into new values corresponding to $\mt$
and $\delta \mt$ inputs different from those employed here.

If the $\Delta a_{\mu}$ discrepancy is real, it points to ``new
physics'', like low-energy supersymmetry. In fact, an intriguing
explanation of $\Delta a_{\mu}$ is provided by some supersymmetric
models, where it is reconciled by the additional contributions of
supersymmetric partners~\cite{NP} and one expects $\mh \lesssim
135$~GeV for the mass of the lightest scalar~\cite{DHHSW}.  If,
instead, the deviation is caused by an incorrect leading-order
hadronic contribution, it leads to a larger $\Delta \alpha_{\rm
  had}^{(5)}(\mz)$ and, correspondingly, to low values of $\mhUB$,
thus leaving a very narrow range for the {\small SM} Higgs boson mass.

\begin{acknowledgments}
\noindent 
We would like to thank G.~Colangelo, G.~Degrassi, S.~Eidelman,
A.~Ferroglia and T.~Teubner for very useful discussions, and
S.~Eidelman, S.~M\"uller, F.~Nguyen and G.~Venanzoni for precious help
with the experimental data of the hadronic cross sections.  M.P.\ also
thanks the Department of Physics of the University of Padova for its
support.
The work of M.P.\ was supported in part by the E.C.\ Research Training
Networks under contracts MRTN-CT-2004-503369 and
MRTN-CT-2006-035505. The work of W.J.M.\ was supported by U.S.\ DOE grant
DE-AC02-76CH00016. The work of A.S.\ was supported in part by the
U.S.\ NSF grant PHY-0245068.

\end{acknowledgments}


\end{document}